\documentstyle[12pt,epsfig]{article}

\newcommand {\vp} {v \!\cdot\! p}

\newcommand {\be} {\begin{equation}}
\newcommand {\ee} {\end{equation}}
\newcommand {\ba} {\begin{eqnarray}}
\newcommand {\ea} {\end{eqnarray}}

\newcommand {\no} {\nonumber}

\def\beq{\begin{eqnarray}}
\def\eeq{\end{eqnarray}}
\def\slash#1{#1 \hskip -0.5em / }

\def\gl#1{(\ref{#1})}

\def\Tr#1{{\rm Tr}\left[#1\right]}

\begin{document}



\thispagestyle{empty}
\begin{titlepage}

\begin{flushright}
HUB--EP--96/40 \\
hep-ph/9608223 \\
August 1996 
\end{flushright}
\vspace{1cm}
\begin{center}
\Large \bf
Extended NJL Model for light and heavy mesons
without $q$-$\bar q$ thresholds
 \\
\end{center}
\vspace{0.5cm}
\renewcommand{\thefootnote}{\fnsymbol{footnote}}
\begin{center}
Dietmar Ebert\footnotemark[1],
Thorsten Feldmann\footnotemark[1]\footnotemark[3], 
\\
{\sl Institut f\"ur Physik, Humboldt--Universit\"at zu Berlin,\\
 Invalidenstra\ss e 110, D--10115 Berlin, Germany}\\
\vspace{0.3cm}
Hugo Reinhardt\footnotemark[2],\\
{\sl Institut f\"ur Theoretische Physik,
      Universit\"at T\"ubingen,\\
   Auf der Morgenstelle 14, D--72076 T\"ubingen, Germany}
\end{center}
\vspace{0.6cm}
\begin{abstract}
\noindent
We consider the NJL model as an effective quark theory
to describe the interaction 
which is responsible for the quark flavor dynamics
at intermediate energies.
In addition to the usual ultraviolet cut-off which is
necessary since the model is non-renormalizable,
we also introduce an infrared cut-off which drops off the
unknown confinement part of the quark interaction, which
is believed to be less important for the flavor dynamics.
The infrared cut-off eliminates all $q$-$\bar q$ thresholds,
which plague the application of the usual NJL model beyond
low-energy pion physics. We apply this two--cut-off prescription
to the extended NJL model with chiral and heavy quark symmetries
proposed recently by us.
We find a
satisfactoring description even of the heavy
mesons with spin/parity $J^P=(0^+,1^+)$.
Furthermore, 
the
shape-parameters of the Isgur-Wise function 
are studied as a function of the residual heavy meson mass.
\end{abstract}
\vspace{3em}


\vspace{1em}


\setcounter{footnote}{1}
\footnotetext{Supported by
{\it Deutsche Forschungsgemeinschaft} under contract Eb 139/1--2.}
\setcounter{footnote}{2}
\footnotetext{Supported by COSY under contract 41170833.}
\setcounter{footnote}{3}
\footnotetext{E-mail: {\sc feldmann@pha2.physik.hu-berlin.de}}
\vfill
\end{titlepage}

\renewcommand{\thefootnote}{\arabic{footnote}}
\setcounter{footnote}{0}
\setcounter{page}{1}

\section{Introduction}

The NJL model has proved quite successful in describing
the low-energy meson physics \cite{EbRe86}.
Its successes stem from the fact that it embodies 
chiral symmetry and its spontaneous breaking.
Unfortunately, the model is not renormalizable.
This necessitates the introduction of an ultraviolet cut-off
$\Lambda$, indicating the range of validity of the model.
Futhermore the model does not confine the quarks, and as
a consequence the model suffers from unphysical $q$-$\bar q$
thresholds, which severely restrict the applications of
the model.
The lack of quark confinement in NJL-type
models is related
to the fact, that the very infrared region of 
the gluon-mediated quark interactions is not
adequately incorporated.
In fact, such models are usually constructed
to describe the low-energy properties of mesons, which
are assumed to be dominated
by the quark flavor dynamics at intermediate scales,
say between a confinement scale of a few hundred MeV
and a scale of about 1~GeV. The success of these
models is related to the very fact, that 
chiral symmetry is dynamically broken just in 
this energy region,
and that the picture of
(constituent) quarks interacting
with mesons may be a reasonable approximation
to strong interactions.
Since on the other hand
this picture must be wrong for very low scales
where confinement becomes definitely important,
one better excludes the infrared region explicitly
by introducing an IR cut-off  as long as one has no definite
suitable way of introducing confinement\footnote{There
are non-local extensions of the NJL model (see e.g.\ \cite{x} and 
refs.\ therein) where
the thresholds are avoided. However, there the beauty
and simplicity of the local NJL model is lost, i.e.\ one
has to deal with integral equations, that must be solved
numerically.}.
We will show that the problems connected with
unphysical quark-antiquark thresholds are then automatically
removed.

As an illustration we apply the two--cut-off prescription
to the recently proposed extended NJL model with both
chiral symmetry for the light quarks $m_q \ll \Lambda_{QCD}$ and
heavy quark symmetries\footnote{see 
e.g.~refs.~\cite{IW,Falk,HQET,DonoghueWise,Casalbuoni}}
for the heavy flavors $m_Q \gg \Lambda_{QCD}$
\cite{we}.
There, the usual low-momentum expansion (LME) 
has been applied to analyze the quark determinant,
and reasonable results have been obtained for nearly
all heavy meson observables.
However, for  
the slope-parameter of the Isgur-Wise function
we found a smaller value than expected, and 
the heavy meson states with
spin/parity $J^P=(0^+,1^+)$ could not be well
described. 
It seems therefore 
necessary to study the influence
of the external momenta more carefully, which
is however not under control
as long as the quark model under concern has
no explicit implementation of confinement. 
In this case, for meson momenta larger than the (constituent)
quark masses, the decay of a meson into a real
quark-antiquark pair
becomes kinematically allowed, and one runs into
trouble with unphysical imaginary parts due to quark-antiquark
thresholds. The 
LME circumvents this problem; in this
way the low-energy theorems in the light meson sector
can be recovered (see e.g.~\cite{EbRe86}), while in
the heavy meson sector the LME seems
to be insufficient since
a smooth dependence on external momenta is not 
always found, i.e.\ the LME does not converge for
momenta beyond thresholds.

We will first present the regularized NJL model
and discuss the physical implications of the infrared 
cut-off on the spontaneous breaking of chiral symmetry.
For this purpose we study the gap-equation that describes
the dynamical generation of the constituent quark mass
in dependence of the infrared cut-off. Then we apply the
regularization method to the extended NJL model with
chiral and heavy quark symmetries \cite{we}.
The mass spectrum of heavy
mesons is found, which indeed develops a 
smooth dependence on external momenta -- even beyond
the apparent thresholds -- and leads
to reasonable predictions.
As an application, we finally discuss the slope and curvature
of the
Isgur-Wise function for finite external momenta
and compare to the phenomenological findings.

\section{Implications of an infrared cut-off in the
	NJL model}

The NJL model rewritten in bosonic degrees of freedom
(see eq.~\gl{eff} below)
leads to an effective meson action where besides 
meson mass terms the quark determinant ${\rm Tr}  \ln \, i\slash D$
contributes. Here $i\slash D$ is the Dirac operator of the
quarks in the presence of meson fields.
The real part of the quark determinant is a diverging
object. Using a proper-time regularization, this quantity
becomes (after continuation to Euclidean space)
\beq
\ln \left| \det i \slash D \right|
&\to&
- \frac{N_c}{2} \, \int_{1/\Lambda^2}^{1/\mu^2} \,
\frac{ds}{s} \, \int \Tr{ e^{-s \, \slash D_E^\dagger  \slash D_E}}
\ .
\label{reg}
\eeq
Here $\Lambda$ is the before-mentioned
ultraviolet cut-off.
Furthermore we have introduced here also an infrared cut-off
$\mu$, which takes into account that our model is not applicable
in the very low-energy regime due to the lack of confinement.
This  infrared cut-off is the new feature of the present
analysis. All previous applications\footnote{An exception is
ref.~\cite{bardeen} where a sharp momentum cut-off 
was used for both the ultraviolet and infrared regime. However,
the authors do not take into account the effect of the
external momenta and consequently do not address the question
of quark-confinement (In fact they use the limit $\mu,m_q,\vp
\ll \Lambda$ for a rough estimate).} of
the NJL model have effectively used the limit $\mu \to 0$.
A simple way to understand that the suggested method indeed
removes the quark-thresholds is to consider the simplest
Greens function obtained from eq.~\gl{reg}, namely the
effective light quark propagator. In Euclidean space one obtains
for the denominator
\beq
\int_{1/\Lambda^2}^{1/\mu^2} \, ds \, \exp\left[ - s (k_E^2+m^2)
\right]
&=&
\frac{e^{- \frac{k_E^2+m^2}{\Lambda^2}}-
	e^{- \frac{k_E^2+m^2}{\Lambda^2}}}{k_E^2 + m^2} \ .
\eeq
Obviously, the former pole at $-k_E^2 = k^2 = m^2$
is removed by the infrared cut-off.
We consider
 $\mu$  as an adjustable parameter which may be
interpreted as the scale 
up to which the quarks have been integrated out.
Especially, at $\mu = \Lambda$ (which is the UV cut-off of the
model) the contribution of the 
quark determinant vanishes by construction, and we recover
the original interaction of (current) quarks with non-dynamical
meson fields,
where chiral symmetry is not (yet) broken.
Note that the non-renormalizable NJL model is defined only
after the regularization method has been specified, and in
this sense the ultraviolet cut-off $\Lambda$ and the infrared
cut-off $\mu$ together with the proper-time regularization
are parts of the definition of the model.

\begin{figure}[hbt]
\begin{center}
\psfig{file = 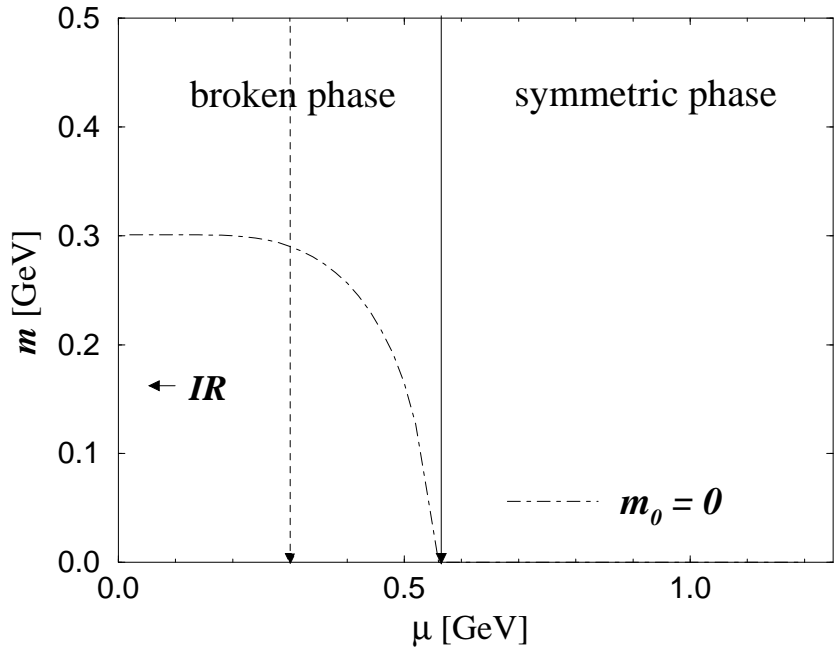, bb = 0 50 580 720, width=10cm, angle = -90}
\end{center}
\caption{The light quark constituent mass $m$ following
from the gap-equation as a function of the scale $\mu$ for
the chiral limit $m_0=0$. The bold vertical line denotes
the separation of the chirally symmetric and chirally broken
phase. 
The dashed
vertical line denotes the chosen
separation of the IR (confinement) region at a
scale $\mu \approx 300$~MeV. 
}
\label{mfig}
\end{figure}

In the NJL model \cite{EbRe86} the spontaneous breaking
of chiral symmetry is realized dynamically
through a non-vanishing VEV for the scalar field $\Sigma$
(chiral radius) which is determined by the following
gap equation
\beq
\langle \Sigma \rangle &=&
m = m_0 + 8 \, m \, G_1 \, I_1(m^2) \ ,
\qquad 
I_1(m^2) = \frac{N_c}{16 \, \pi^2} \, m^2 \,
\Gamma\left[-1,m^2/\Lambda^2,m^2/\mu^2\right] \ ,
\label{gap}
\eeq
This equation defines the constituent quark mass $m$ as an
implicit 
function of the scale $\mu$ for given values
of the remaining model parameters.
Here $\Gamma\left[ \alpha,x,y\right] =
\int_x^y dt \, \exp(-t) \, t^{\alpha-1}$ denotes
the incomplete Gamma function.

For a numerical discussion we use
the coupling constant and the cutoff 
from ref.~\cite{we} as
 $G_1 =5.25$~GeV$^{-2}$ and $\Lambda=1.25$~GeV.
Fig.~1 shows the solution of the gap equation
$m$ as a function of the IR scale $\mu$ in the chiral limit
($m_0 = 0$).
For scales $\mu < 0.3$~GeV the order parameter $m$ of 
the chiral symmetry breaking is almost independent of
$\mu$, while it decreases rapidly when $\mu$ exceeds
a value of $\mu \approx 0.3$~GeV and vanishes at
a critical value $\mu_c$ given by
\beq
\mu^2_c &=&  \Lambda^2 - \frac{2 \, \pi^2}{3 \, G_1} 
	\approx (\mbox{550~MeV})^2 \ .
\eeq
For $\mu > \mu_c$ chiral symmetry is restored.
Consequently, in order to
include the effect of the spontaneous breaking of chiral symmetry,
it is sufficient to consider the quark flavor dynamics 
down to scales of about $\mu \approx 300$~MeV.
At lower values of $\mu$ one expects essential effects
of the IR singular gluon propagator, leading to a
modified gap- (or in this case Schwinger-Dyson-) equation
and a presumably confined quark propagator. Clearly, this
question cannot be addressed within the NJL model and
is outside the scope of the present paper.

\section{Extended NJL model}

In \cite{we} we have presented an
extension of the NJL model which combines
chiral symmetry for light quarks
with the heavy quark symmetries. 
In the heavy mass limit, heavy meson fields
are organized in spin symmetry doublets \cite{Falk} 
\beq
H &=& \frac{1+\slash{v}}{2} \,
      \left\{ i \Phi^5 \gamma_5 + \Phi^\mu \gamma_\mu \right\}
\quad , \qquad v_\mu \Phi^\mu = 0 \ ,\no \\
K &=& \frac{1+\slash{v}}{2} \,
      \left\{ \Phi  + i {\Phi^{5'}}^\mu \gamma_\mu \gamma_5 \right\}
\quad , \qquad v_\mu {\Phi^{5'}}^\mu = 0 \ ,
\eeq
with $v^\mu$ being the heavy quark velocity.
Here $\Phi^5$ and $\Phi^\mu$ denote the pseudoscalar
and vector field of the $J^P = (0^-,1^-)$ multiplet,
$\Phi$ and ${\Phi^{5'}}^\mu$ are the heavy
scalar and axial vector field of the $J^P = (0^+,1^+)$ multiplet.

After introducing light and heavy
meson fields, the
lagrangian of the extended NJL model
can be rewritten as \cite{we}
\beq
{\cal L} & =
& -iN_c {\rm Tr} \ln i\slash{D}^{reg}
-\frac{1}{4G_1} {\rm tr_F}
 \left[ \Sigma^2 -
 \widehat{m}_0 (\xi \Sigma \xi + \xi^\dagger \Sigma \xi^\dagger)
 \right]
\nonumber\\
&&+\frac{1}{4G_2} {\rm tr_F}
 \left[ (V_\mu - {\cal V}_\mu^\pi)^2
       +(A_\mu - {\cal A}_\mu^\pi)^2
 \right]
- \frac{1}{2G_3}
 {\rm Tr} \left[ \slash v \, 
	(\overline{H}+\overline{K}) \, (H + K) \right]
\ ,
\label{eff}
\end{eqnarray}
with $i \slash D$ being the light quark determinant:
\beq
i \slash{D} = i \slash{\partial}
- \Sigma
        + \slash{V} + \slash{A}\gamma_5
        - (\overline{H}+\overline{K})\, (i v \!\cdot\! \partial)^{-1}
              (H + K)
\quad
\label{Dirac}
\eeq
which contains the
light scalar, vector and axial-vector meson fields
$(\Sigma , V_\mu , A_\mu )$ and heavy meson fields $(H,K)$.
$N_c = 3$ is the color factor, $\hat{m}_0$ is the current mass
matrix of light quark flavors and $G_1$, $G_2$, $G_3$ are
independent coupling constants of four--quark interactions allowed
by chiral and heavy quark symmetries (dim~$G_{1,2,3}=$~mass$^{-2}$).
For the light octet of Goldstone bosons we use the common
non--linear representation
$\xi = \exp (i \pi / F)$
where $\pi = \pi^a \lambda_F^a/2$ and $F$ is the bare
decay constant. 
The induced (axial-)vecotr fields
$ {\cal V}_\mu^\pi = i/2 (\xi \partial_\mu \xi^\dagger +
             \xi^\dagger\partial_\mu \xi)$,
$ {\cal A}_\mu^\pi = i/2 (\xi \partial_\mu \xi^\dagger -
             \xi^\dagger\partial_\mu \xi)$
arise from a chiral rotation of the quark fields.
Finally, the light scalar field $\Sigma$ achieves a non--vanishing
vacuum expectation value indicating the spontaneous breaking of
chiral symmetry (see eq.~\gl{gap} above). For further
details we refer the reader to ref.~\cite{we}.

For heavy mesons with the residual external momentum
$\vp$ the
self-energy contribution from the quark loop,
together with the constant terms in eq.\ (\ref{eff}),
determines the residual masses $\Delta M = M - m_Q$ and
the renormalization factors $Z$
\beq
 \frac{1}{ 2 \, G_3} &\stackrel{!}{=}& 
\left. 
I_3(\vp,m^2) \ \left(\vp \pm m \right) + I_1(m^2)
\right|_{\vp=\Delta M_{H,K}} \ ,
\label{selfcon}
\label{A}
\eeq
\beq
Z_{H,K} = \left. \left( I_3 (\vp,m^2) +
\frac{\partial}{\partial \, \vp} \, I_3(\vp,m^2) \
(\vp \pm m)  \right)^{-1} 
\, \right|_{\vp=\Delta M_{H,K}}
\ ,
\label{Zfac}
\eeq
where the upper sign $(+)$ is for the $H$-fields and
the lower sign $(-)$ for the $K$-fields, indicating
the mass splitting of the two parity conjugated states
due to the broken chiral symmetry for $m \neq 0$.
The function $I_3$ which diverges linearly with the UV cutoff
$\Lambda$ is represented by the proper-time integral
\beq
I_3(\vp,m^2)&=&
\frac{N_c}{16\pi^2} \, \int_{1/\Lambda^2}^{1/\mu^2}
\frac{\sqrt{\pi}\,ds}{s^{3/2}} \,
\left( 1 + {\rm Erf}\left[\sqrt{s}\,\vp\right] \right) \,
\exp(- s (m^2 - \vp^2)) \ ,
\eeq
with 
$
{\rm Erf}\left[a\right] =
\frac{2}{\sqrt{\pi}} \int_0^a dx \, \exp(-x^2)
$
being the Gaussian error function.

Finally, the Isgur-Wise function is calculated from
the quark determinant by inserting an arbitrary
current for a weak transition between heavy 
quarks of different velocity $\bar Q'_{v'} \, \Gamma \, Q_v$
with the result 
\beq
\xi(\omega) &=&
\left. Z_H \, \left( \frac{2}{1+\omega} \, (I_3 + \vp \, I_5(\omega)) 
	\ + \ m \, I_5(\omega) \right)
\right|_{\vp = \Delta M_H}
\label{B}
\ .
\eeq
The function $I_5$ of the momentum transfer
$\omega = v\!\cdot\!v'$ is expressed through the parameter
integral
\beq
I_5(\omega)
& = & \int_0^1 dy \,
\frac{\partial}{\partial \vp_y} \,
\frac{ I_3(\vp_y, m^2) }
 {1 + 2 y (1-y)(\omega - 1)}
\ , \qquad {\vp}_y \equiv 
\vp \, / \,\sqrt{1+ 2 y(1-y) (\omega -1)} \ .
\eeq
The normalization condition $\xi(1)=1$ comes out naturally
as a Ward identity relating \gl{A} and \gl{B}.

In ref.~\cite{we} we have discussed the Isgur-Wise function
in the LME, which refers to the special case $\vp = 0$.
In this case one has $I_5(\omega) = 2 I_2 \, r(\omega)$ with 
$$
r(\omega) = \frac{\ln(\omega + \sqrt{\omega^2-1})}{\sqrt{\omega^2-1}}
\ .
$$
However, due to the introduced IR cut-off $\mu$ for the
quark determinant \gl{reg} we can give up the LME around $\vp=0$ 
and use eq.~\gl{B} for any value of $\vp$.

Concerning the mass equation
for the heavy mesons (\ref{selfcon}),
let us point out that
for $\vp > m$ the integral $I_3$ generates
non-physical behaviour in the IR region ( $\mu \to 0$, $s\to \infty$).
This is easily understood, since through the optical
theorem this is connected with
an imaginary part reflecting a decay of a meson into
real quarks which should be absent in a confining theory. 
For practical purposes for sufficiently small values $\vp < m$,
an expansion in $\vp$ (LME) might be sufficient
\cite{we}.
For the general case one better retains a finite value of $\mu$
which (in order to be on the safe side)
we will chose in the following as
$\mu=300$~MeV.
The basic assumption of chiral quark models like the one
presented here is then
to assume that the 
IR region of
the quark dynamics excluded by the cut-off $\mu$ 
might be unimportant for the
low-energy properties of mesons.

\section{Heavy meson spectrum and Isgur-Wise function}
 
\begin{figure}
\begin{center}
\psfig{file = 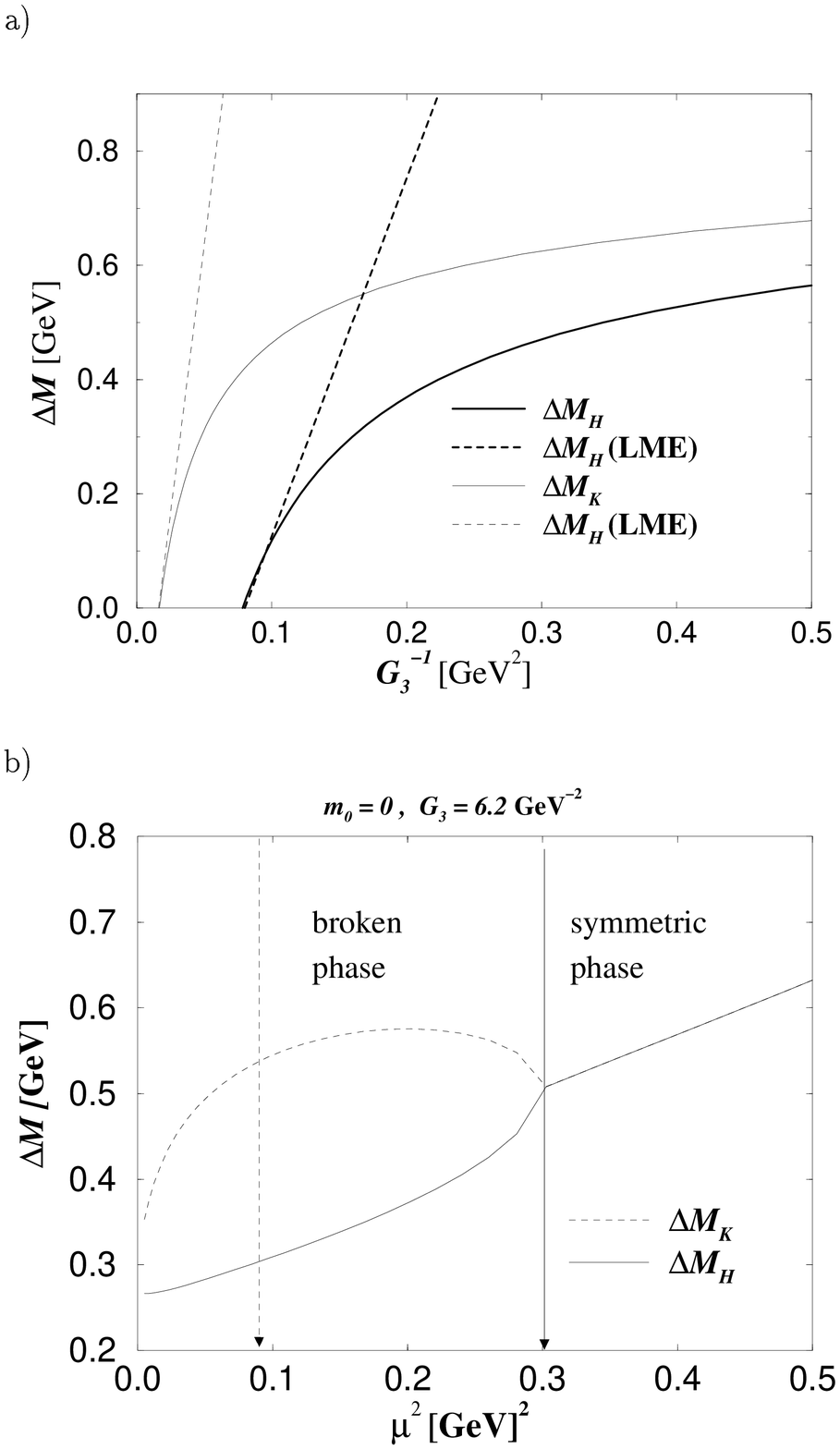, bb = 0 50  580 720, width=15cm}
\end{center}
\caption[]{
a) The behaviour of the residual masses $\Delta M_H$ (bold
line) and  of $\Delta M_K$ (thin line)
as a function of $1/G_3$ from
eq.~(\ref{selfcon}) for an infrared cut-off $\mu=300$~MeV
compared to the LME with $\mu = 0$
(dotted lines).\\
b) The behaviour of $\Delta M_H$ and $\Delta M_K$
in the chiral limit $m_0 \to 0$ 
as a function of $\mu$ for a fixed value of $G_3=6.2$~GeV$^{-2}$.
Due to the restoration of chiral symmetry for $\mu\geq\mu_c$
the heavy meson fields $H,K$ fall into spin/parity quartets. 
}
\label{f1}
\end{figure}

Through eq.\ (\ref{selfcon}) the coupling
constant $G_3$ is related to the
heavy meson mass spectrum. For the light quark masses
we will use the values $m_u=m_d=300$~MeV and $m_s=510$~MeV.
Let us first compare the numerical results for
eq.\ (\ref{selfcon}) in contrast to the LME with $\mu =0$ 
in Fig.~2a.
We observe 
significant differences between the
two approaches for residual meson masses $\Delta M > m$.
Furthermore, we can study the $\mu$-dependence of the
heavy meson masses (see Fig.~2b). For $\mu\sim300$~MeV
we observe a reasonable  mass-splitting
between the parity conjugated fields $H,K$ of about
200--300~MeV. 


In
Table~\ref{t2} we present the results in comparison
with the LME,
including the model results for
the weak decay constant of the heavy meson
given as \cite{we}
$
\sqrt{M_H} \, f_H = \sqrt{Z_H}/{G_3} \ .
$

\begin{table}[hbt]
\begin{center}
\begin{tabular}{l|l|cc|cc|cc}
&$G_3$ [GeV]$^{-2}$ &
$\Delta M_H^u$ & $\Delta M_H^s $ &
$\Delta M_K^u$ & $\Delta M_K^s $ &
$f_B$ & $f_B^s/f_B$
\\
\hline
&8.2 & 200 & 255 & 500 & 685 & 135 & 1.04\\
$\mu = 300$~MeV &6.2 & 300 & 360 & 545 & 715 & 140 & 1.04\\
&4.5 & 400 & 475 & 590 & 750 & 150 & 1.04 \\
\hline
LME \protect\cite{we} & 8.7 & 220 & 320 & -- & -- & 180 & 1.1 
\end{tabular}
\end{center}
\caption{Several heavy meson parameters in the extended NJL model (masses
and decay constants in [MeV]).} 
\label{t2}
\end{table}

Compared to the LME,
the results for the heavy meson spectrum
are improved, in the sense that they are
leading to realistic masses
even for the states with $J^P=(0^+,1^+)$, while
the results for $J^P = (0^-,1^-)$ are -- as expected --
only slightly effected.
Note that all masses are now rather smooth functions
of the coupling constant, which was not the case for
the LME with $\mu$=0. 
The values for the weak decay constant lie on the lower
side of the range obtained from lattice QCD \cite{wittig},
$f_B = 170 \pm 55$~GeV.
The light flavor dependence can be studied by considering
the $SU(3)_F$ mass-splitting, which in our case is induced
by different light quark masses $m_u \neq m_s$.
Experimentally one has
$
M_{B_s} - M_{B_u}
\simeq M_{D_s} - M_{D_u} \simeq 
 100\mbox{~MeV}~\cite{PD}\ .
$
Our model predicts somewhat smaller values 
$\Delta M_{H_s} - \Delta M_{H_u} \approx 55 - 75 ~\mbox{MeV} \ .
$ 
Note that this quantity only receives small $1/m_Q$ corrections
of the order $(m_s-m_u)/m_Q$.

Let us now concentrate on the Isgur-Wise function.
In ref.~\cite{Neubert}
long-distance ($1/m_Q$) and short-distance ($\alpha_s(m_Q)$)
corrections have been estimated,
which relate the Isgur-Wise function $\xi(\omega)$
to
the experimentally accessible form factor(s)
${\cal F}(q^2)$.
From this one of the CKM matrix elements has been
estimated as $|V_{cb}|=0.039 \pm 0.002$.
In particular, 
a phenomenological
range for the slope parameter $\rho = \sqrt{-\xi'(1)}$
is quoted, $0.84 \leq \rho \leq 1.00$,
and a correlation between the slope and the curvature
$c_0 = \xi''(1)/2$ is estimated as
$c_0 \simeq 0.72 \, \rho^2 - 0.09$ \cite{caprini}.

The result of the NJL model gives
for different values of $\vp = \Delta M_H$
\beq
\rho |_{\vp =0} &=& 0.67 \ \cite{we} \no \\
\rho |_{\vp =\mbox{\footnotesize 200~MeV}} &=& 0.77 \ , \quad  c_0 = 0.38  \ ; \no \\
\rho |_{\vp =\mbox{\footnotesize 300~MeV}} &=& 0.84 \ , \quad  c_0 = 0.49  \ ; \no \\
\rho |_{\vp =\mbox{\footnotesize 400~MeV}} &=& 0.94 \ , \quad  c_0 = 0.72  
\ .
\eeq
The dependence on the external momentum
is obvious. Values between $\Delta M_H = 300$~MeV
and 400~MeV
reproduce the phenomenological situation, and the
values of $\Delta M_H$, $\rho$ and $c_0$ are positively correlated.
Note that the inclusion of a finite curvature for
the Isgur-Wise form factor is important 
in order to obtain an accurate fit to the data
for the $B \to D \ell \nu$ decays which were used
for the extraction of $|V_{cb}|$ \cite{jimack}.

\begin{figure}
\begin{center}
\psfig{file = 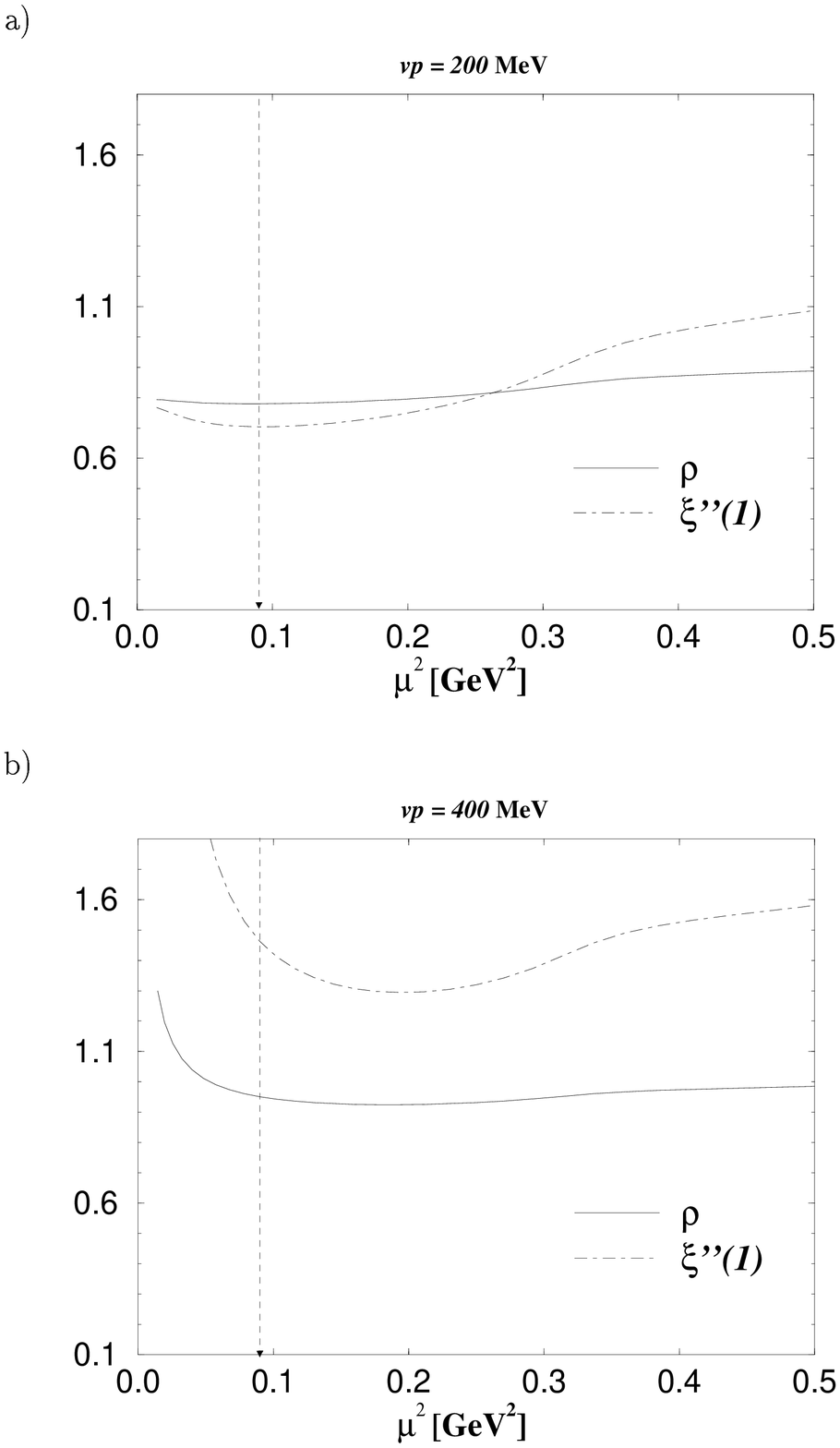, 
	bb = 0 50  580 720, width=15cm}
\end{center}
\caption{
The behaviour of the shape parameters of 
the Isgur-Wise function, $\rho$ and $\xi''(1)$ as 
functions of the IR scale $\mu^2$ for a) $\vp=200$~MeV and
b) $\vp=400$~MeV. The vertical line denotes the choice
$\mu = 300$~MeV.} 
\label{figiwmu}
\end{figure}

Finally, we like to investigate the $\mu$-dependence of
the shape parameters $\rho$ and $\xi''(1)$ which
is of course rather stable for $\vp \leq m$.
For $\vp > m$, our choice for the
IR cut-off scale is justified,
since for $\mu < 300$~MeV the predictions for
$\rho$ and $c_0$ become unreliable.
This is illustrated in Fig.~3 for the two
cases $\vp = 200$~MeV and $\vp = 400$~MeV, respectively.

\section{Summary}

In this letter we have shown 
how the NJL model, which is phenomenologically successful
at low meson energies, can be extended to higher energies.
In particular we have shown that the appearance of quark
thresholds, which plagues the application of the NJL model
to higher external momenta can be avoided.
This has been accomplished by introducing
a finite adjustable IR scale when integrating out
the quarks. 
We have found that for the spontaneous breaking
of chiral symmetry the momentum regime between
$\mu = 300$~MeV and $\Lambda \approx 1.25$~GeV is responsible.
Retaining a finite
value for the IR scale (the choice $\mu \approx 300$~MeV has been
favored by the data) 
prevents the quark determinant
to suffer from unphysical quark-antiquark thresholds, which
would otherwise occur in chiral quark models without explicit quark
confinement.

Including heavy quarks in the infinite mass limit
one then obtains a smooth dependence on the external
momenta for the heavy meson observables,
and a reasonable mass spectrum is found even for
the heavy $J^P = (0^+,1^+)$ spin-symmetry multiplet.
For the weak decay constants we found values of the
order of $f_B = 140$~MeV.

For the shape parameters of the Isgur-Wise function
we have found a non-negligible dependence on external momenta,
such that the value of the slope parameter $\rho$ is
positively correlated to the value of $\Delta M_H$.
For $\Delta M_H =300 (400)$~MeV we found $\rho = 0.84 (0.94)$
which is consistent with the experimental and theoretical situation.
We also give an estimate for the curvature at the non-recoil point,
$c_0 = \xi''(1)/2 \approx 0.49 (0.72)$. 
At intermediate scales $\mu > 300$~MeV
the shape parameters develop only a weak dependence on $\mu$.

In future work our analysis can be helpful also
for the investigation of higher
resonances, like the heavy $J^P = (1^+,2^+)$ multiplet,
within NJL-type quark models,
where the dependence on external momenta is assumed to
be even more important.
Also the description of the higher excited light flavor
states within the NJL model might be considerably improved.
It can also be interesting to compare the approach
presented here, with the results obtained
in other relativistic quark models, which explicitly
include confinement in a heuristic way.

\end{document}